\documentclass{elsart}
\usepackage{graphicx}
\usepackage{hyperref}
\begin{document}
\begin{frontmatter}
\journal{Surface Science}
\title{Tight binding studies of strained Ge/Si(001) growth}
\author{K.~Li\thanksref{china}},
\ead{khli@xmu.edu.cn}
\thanks[china]{Permanent Address: Physics Department, Xiamen University, Xiamen, 361005, P.R.China}
\author{D.R.~Bowler\corauthref{drb}} and 
\ead{david.bowler@ucl.ac.uk}
\ead[url]{http://www.cmmp.ucl.ac.uk/$\sim$drb/research.html}
\corauth[drb]{Corresponding author}
\author{M.J.~Gillan}
\ead{m.gillan@ucl.ac.uk}
\address{Department of Physics and Astronomy, University College London,
Gower Street, London WC1E 6BT, UK}

\begin{abstract}
Experimental observations of the growth of more than one monolayer of
Ge on Si(001) show a progression of effects beyond the $(2 \times N)$
reconstruction which is seen at submonolayer coverages: a reduction in
the value of \textit{N} with coverage; formation of straight trenches
of missing dimer vacancies; and formation of the $(M \times N)$
``patch'' structure. We present tight binding calculations which
investigate the energetics and geometries associated with these
effects, and extend our earlier treatment of formation energies for
reconstructions with different stoichiometries to the case of several
layers of Ge.  The results provide explanations for the various
effects seen, and are in good agreement with experimental
observations.
%
\end{abstract}

\end{frontmatter}

\section{\label{sec:intro}Introduction}
Understanding the relationship between strain and the formation of
nanostructures is extremely important as practical nanoelectronic
devices are sought.  The growth of Ge on Si(001) is particularly
interesting in this context, not only because of the small, three
dimensional features which form at high coverage (the so-called ``hut
clusters''\cite{Mo1991}) but also because of the direct compatibility
with standard group IV semiconductor technology.  In this paper, we
present tight binding calculations of various aspects of Ge growth on
Si(001) for coverages between one and three monolayers (ML) of Ge.
This work follows naturally from our previous paper\cite{Oviedo2002},
where we studied the $(2 \times N)$ reconstruction which forms at
sub-monolayer coverages using first principles electronic structure
techniques.

\begin{figure}
\includegraphics[width=\columnwidth,clip]{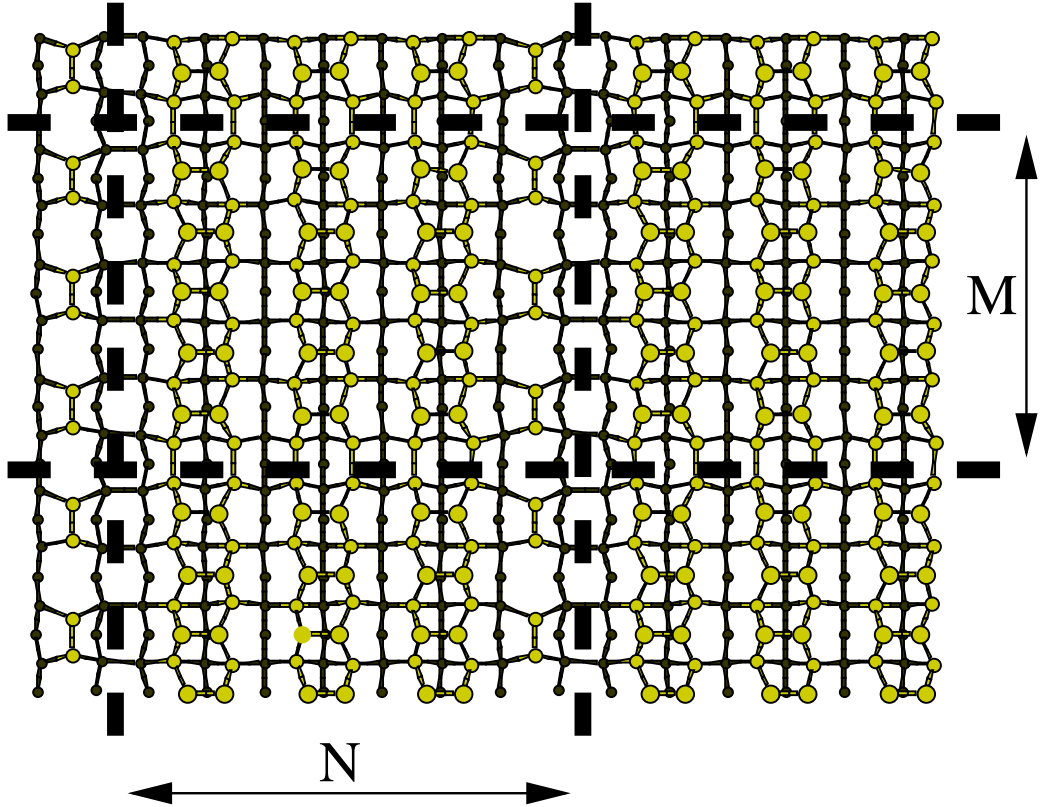}
\caption{\label{fig:IllusMbyN}A top-down view of the $(M \times N)$
surface for two layers of Ge on Si (only the top four layers are shown
for simplicity; the Ge is shown as light circles, with the top layer
larger than the second layer, while Si is shown as dark circles).
The underlying $(2 \times N)$ reconstruction has dimer rows running
across the page horizontally, while the top-layer $(M \times 2)$
reconstruction has dimer rows running vertically.  For this cell,
$N=8$ (equivalent to three dimer rows and a missing row in the top
layer) and $M=6$ (equivalent to five dimers and a missing dimer in the
top layer).}
\end{figure}

As the coverage of Ge on Si(001) approaches 1ML, it forms the
well-known $(2 \times N)$ reconstruction, where regularly spaced
missing dimers appear in the surface; experimental observations of
\textit{N} show that it typically lies between 8 and 12, with the
spacing dependent on growth source (gas source, e.g. GeH$_4$, or solid
source) and conditions (growth rate and
temperature)\cite{Goldfarb1997,Voigt1999}.  As the thickness of the
deposited layer increases, various further effects are seen on this
reconstruction, again depending on growth source and conditions for
their onset: the value of \textit{N}
decreases\cite{Voigt1999,Wu1995,Huang1997}; the missing dimers align
in neighbouring dimer rows to form straight
trenches\cite{Wu1995,Huang1997}; a new, patch-like reconstruction with
a double periodicity (called the $(M \times N)$ reconstruction) forms
(illustrated in
Fig.~\ref{fig:IllusMbyN})\cite{Goldfarb1997,Kohler1992,Tomitori1994};
the roughness of step edges changes (with the different step types
eventually becoming equally rough, leading to steps running along the
elastically soft (100) and (010) directions)\cite{Wu1995}; and finally
three dimensional structures such as ``hut'' pits and clusters
form\cite{Mo1991,Goldfarb1997b}.  We shall investigate the first three
of these phenomena in this paper.

The modelling of the formation of straight trenches and the $(M \times
N)$ reconstruction requires large unit cells, while the comparisons we
are performing require a quantum mechanical technique for accuracy
(for instance, to correctly model the dimer buckling on the surface).
We have chosen to use an $\mathcal{O}(N)$ (or linear scaling) tight
binding technique to meet these criteria\cite{Goedecker1999} (this is
a technique where the computer effort required scales linearly with
the system size at the expense of an approximation, rather than with
the cube of the system size, as is the case for standard techniques,
referred to as ``diagonalisation'').  However, this introduces two
levels of approximation: the first, inherent in the tight binding
parameterisation; the second, in the use of an ${\mathcal O}(N)$
method.  We test these approximations separately.  First, we test the
parameterisation by comparing the results of exact diagonalisation to
\textit{ab initio} results.  Second, we check the accuracy of the
${\mathcal O}(N)$ technique against the exact diagonalisation tight
binding results.  This gives us a clear idea of what errors are
introduced and where.  Starting from the $(2 \times N)$
reconstruction, we investigate first the dependence of the value of
\textit{N} on the depth of the Ge layers.  We then calculate the
energetics of kinks in trenches of missing dimers with increasing Ge
depth, in order to understand the formation of straight trenches of
missing dimers.  Finally we address the $(M \times N)$ reconstruction,
calculating minimum energy values of \textit{M} for a fixed value of
\textit{N}.  In all of the above calculations, we will also address
the question of an appropriate energy to use for Ge dimers when
comparing reconstructions of different stoichiometry, which is closely
related to the problem of finding an appropriate chemical potential
for Ge.

The previous work in this area only addressed the $(2 \times N)$
reconstruction, and investigated the dependence of \textit{N} on depth
of Ge using empirical potentials, such as the Stillinger-Weber and
modified Keating forms~\cite{Tersoff1991,Tersoff1992,Liu1996}.  These
studies found results in broad agreement with experiment, and our
previous work\cite{Oviedo2002} provides support for the mechanisms
suggested by these empirical studies.  We note that intermixing of Si
and Ge may well occur during experimental
growth\cite{Voigt1999,Huang1997}, though this depends on experimental
conditions, and is thought to be largely suppressed after one
monolayer of Ge.  The likely effect would be to reduce the surface
strain somewhat; we have chosen not to account for intermixing for
three reasons.  First, there are experimental results available for
situations where intermixing is almost entirely suppressed, with which
we can make direct contact.  Second, it adds significant additional
complication to the problem, where we are trying to find the
underlying mechanisms. Third, the effects are likely to be relatively
small.  Indeed, in previous empirical work\cite{Liu1996}, the authors
concluded that the Ge/Si interlayer mixing could not be substantial.
In a recent detailed \textit{ab initio} study of
intermixing\cite{Uberuaga2000}, it was shown that, while Ge can
penetrate to the fourth layer, this is likely to be at the level of a
few percent at a typical low growth temperature (e.g. 500$^\circ$C).
Hence, we neglect this effect in the present work.

The rest of the paper proceeds as follows: in the next section we
review the details of the computational techniques used, consider the
question of what energies to use when comparing reconstructions of
different periodicity (and hence stoichiometry) and test the accuracy
of our methods; after this, we present our results for the different
situations considered (the $(2 \times N)$ reconstruction with
increasing Ge depth; trench kinking with increasing Ge depth; the $(M
\times N)$ reconstruction), and we finish with a discussion of the
results.

\section{\label{sec:details}Computational Details}

\subsection{Tight Binding Parameters}
The tight binding technique has been discussed elsewhere in
detail\cite{Goringe1997}, as have the ideas and implementation of
${\mathcal O}(N)$ techniques\cite{Goedecker1999}.  We used a
parameterisation for Si-Si bonds (and Si-H bonds for termination)
specifically designed for and tested extensively on the Si(001)
surface\cite{Bowler1998}, and a parameterisation for Ge-Ge and Ge-Si
bonds again tested on the Si(001) surface, and fitted to reproduce the
experimental Ge lattice constant (giving a lattice mismatch of 4\%
between Ge and Si)\cite{Bowler2002}.  We used the \textsc{Oxon} code
to perform exact diagonalisations, and the \textsc{DensEl} code (an
implementation\cite{Goringe1995} of the auxiliary density matrix (ADM)
technique\cite{Li1993}) for linear scaling calculations.

We work with periodic boundary conditions, simulating a surface via a
slab with a vacuum gap equivalent to eight atomic layers of Si.  In
all cases, the slab modelled was ten atomic layers deep, with the
bottom layer terminated in hydrogen.  The length and width used
depended on the calculations: for the $(2 \times N)$ calculations,
cells were one dimer row wide, and ranged from four to twelve dimers
long (from 94 to 286 atoms); for trench kinking calculations, they
were between two and six dimer rows wide and eight (1 or 2ML Ge) or
six (3ML Ge) dimers long (from 284 atoms to 1,140 atoms); and for the
$(M \times N)$ reconstructions, they were typically six dimer rows
wide and eight dimers long (around 1,100 atoms).  We investigated the
importance of substrate relaxation, and found that in most cases (in
all cases with 2 or more monolayers of Ge on Si(001)) allowing four
layers of silicon below the Ge to relax gave energy convergence to
within a few meV (and presented the same amount of substrate
relaxation below the Ge).  For the detailed comparison of one
monolayer of Ge on Si(001) presented below, we found that allowing
seven layers of silicon to relax was important.  The reason for this
lies in the rebonding of the second layer atoms across the gap where
the dimer is missing.  When these atoms are silicon, they are
stretched across the gap, and require significant depth for relaxation
of strain.  When they are germanium, the rebonding helps to alleviate
the mismatch with the silicon substrate, and less depth is required.

When using exact diagonalisation to test the parameterisation
involving Ge against earlier DFT results, we used a $4 \times 2 \times
1$ Monkhorst-Pack \textbf{k}-point mesh\cite{Monkhorst1976}, which we
found converged total energies to within 2 or 3 meV.  When performing
the linear scaling calculations, we had to choose a cutoff on the
density matrix, which in the \textsc{DensEl} code is typically
specified as a number of hamiltonian-range hops.  Previous work has
shown that, for semiconductors and insulators, forces and energy
differences are well converged at 3 hops (roughly 6 \AA\ in Si) and
completely converged at 5 hops (roughly 10 \AA\ in
Si)\cite{Bowler1997}.  We present detailed calculations for the energy
of one monolayer of Ge in the $(2 \times N)$ reconstruction below (in
section~\ref{sec:approx}), but the conclusion is that 3 hops is an
appropriate compromise between accuracy and CPU time.

\subsection{\label{sec:Gemu}Comparing different stoichiometries}

In our previous paper\cite{Oviedo2002}, where we presented \textit{ab
initio} calculations of the $(2 \times N)$ reconstruction for
sub-monolayer coverages of Ge, we gave detailed arguments for the
energy to be used for Ge dimers when comparing reconstructions of
different periodicity and hence stoichiometry.  This energy (which is
closely related to a chemical potential) is extremely important when
making contact with the experimentally observed values of \textit{N}
and \textit{M}, which is one of the aims of this paper.  We will
briefly recall the statistical mechanical arguments used before, and
then extend them to the rather different situations found in this
work.

We are assuming that the surface is in thermal equilibrium, which
means of course that we are ignoring kinetic effects --- an assumption
that will be discussed in Sec.~\ref{sec:conc}.  Then, for a given
number of Ge atoms on the surface, the probability of finding them in
a particular arrangement $\Gamma$ will be proportional to $\exp
(-E_\Gamma / k_{\mathrm{B}}T)$, with $E_\Gamma$ the energy for the
arrangement $\Gamma$.

We only need to know how $E_\Gamma$ varies from one arrangement to
another since we are considering a \textit{fixed} number of
atoms. However, it will be helpful to use one specific arrangement as
a ``reference'' arrangement; we choose this to be all Ge atoms paired
into dimers and arranged in a perfect, lattice-matched monolayer
covering a certain area of the surface.  The shape is unimportant, but
it is easier to consider a rectangle.  We now consider forming regular
arrays of missing dimer trenches by removing dimers from the monolayer
and replacing them at the edges of the island, and allowing the system
to relax.  It is convenient to split this process into two parts: (i)
fetch an appropriate number of Ge dimers (equivalent to the number to
be removed) from infinity and place them at the boundary of the layer;
(ii) remove the appropriate Ge dimers to form missing dimer trenches
and relax the system.  We now calculate the quantity $\zeta(N)$, which
is the energy change per Ge dimer in going from a perfect layer to a
periodically reconstructed one.  It is given by\cite{Oviedo2002}:

\begin{equation}
\zeta ( N ) = {( E_{\mathrm{f}} ( N ) + E_{\mathrm{p}} ) \over ( N - 1 )},
\label{eq:zeta}
\end{equation}
where $E_{\mathrm{f}}(N)$ is the formation energy of fully relaxed
missing dimers in the top layer with spacing \textit{N} (defined as
the energy difference between a relaxed surface with a missing dimer
reconstruction and a surface with a perfect top Ge layer), and
$E_{\mathrm{p}}$ is the energy for a Ge dimer in the reference system
(a perfect layer --- the energy is defined as the difference per Ge
dimer between a surface with a perfect top Ge layer, and the surface
formed by removing the top layer of Ge and reconstructing and
relaxing).  By finding the minimum value of $\zeta(N)$ while varying
\textit{N}, we can find the expected value of \textit{N}.  It should
be noted that if $\zeta(N)$ is \textit{positive}, then the
reconstruction is \textit{less} stable than a perfect monolayer of Ge
(which was chosen as the reference system).

Now we consider how to generalise this argument to the systems considered
in this paper.  There are four comparisons that we will make:
\begin{enumerate}
\item The $(2 \times N)$ reconstruction, on multiple layers of Ge
(compared to experiment) (Sec.~\ref{sec:2byNDeep})
\item The energy of missing dimer trenches with and without disorder
(compared to each other) (Sec.~\ref{sec:kinking})
\item The $(M \times N)$ reconstruction, with multiple layers of Ge
(compared to experiment) (Sec.~\ref{sec:mbyn})
\item The energies of $(2 \times N)$ and $(M \times N)$
reconstructions on the same number of layers of Ge (compared to each
other) (Sec.~\ref{sec:Compare1D2D})
\end{enumerate}

For the second of these (the energetics of disorder in missing dimer
trenches), we do not need to concern ourselves with compensation for
different amounts of Ge, as there is no variation in period, just in
arrangements of missing dimers.  For the other three, we need to
identify appropriate energies to use when making the comparisons
discussed above, as well as the physical situation we are dealing
with.

\paragraph*{The $(2 \times N)$ reconstruction on multiple layers of Ge}
This is possibly the simplest case to deal with.  To be clear, we have
a system composed of a Si substrate, with $\theta$ layers of Ge on top,
only the top layer of which has a missing dimer reconstruction (in
other words we have $\theta-1$ full layers of Ge on top of Si as our
substrate).  Then we use the same argument as we used for
sub-monolayer coverages of Ge on Si(001) --- i.e. we consider a pool
of Ge on top of $\theta-1$ full layers of Ge on Si, and examine
different arrangements that we could make with it, calculating
$\zeta(N)$ for each.  The energy that we require for comparisons is
then the energy difference per Ge dimer between $\theta$ full layers
of Ge on Si, and $\theta-1$ full layers of Ge on Si, which we call
$E_{\mathrm{p}}(\theta)$, where $\theta$ is the coverage (or number of
layers).

\paragraph*{The $(M \times N)$ reconstruction on multiple layers of Ge}
The easiest energy to use for this case is little different to that
used above; we assume that we have $\theta-2$ full layers of Ge on top
of Si, and then a further layer of Ge with a $(2 \times N)$
reconstruction (the $\theta-1^\mathrm{th}$ layer).  This acts as our
substrate; since there is a $(2 \times N)$ reconstruction, and the
dimer row direction rotates by $90^\circ$ with each layer, we must form
finite width strips of Ge on this substrate (where the width is
$N/2-1$ dimer rows) to make an $(M \times N)$ reconstruction.  We now
consider our pool of Ge as divided into strips, and make different
arrangements with that, now calculating $\zeta(M)$ for each.  The
energy required is then just the energy difference per Ge dimer
between perfect strips of Ge on a $(2 \times N)$ reconstructed surface
with $\theta-1$ layers of Ge and the $(2 \times N)$ surface
reconstructed surface with $\theta-1$ layers of Ge, which we call
$E_{\mathrm{p}_\mathrm{strip}}(\theta)$, where again $\theta$ is the
coverage (or number of layers).

\paragraph*{The $(2 \times N)$ and $(M \times N)$ reconstructions on multiple layers of Ge}
It is easiest to compare these reconstructions by comparing their
stabilities relative to the perfect surface, which requires
compensation for the missing Ge.  For the $(2 \times N)$
reconstruction, this requires only $E_{\mathrm{p}}(\theta)$, which has
already been calculated above.  For the $(M \times N)$ reconstruction,
we need to perform a little more work, but only using quantities that
we already have --- the energy of Ge dimers in a perfect layer of a
given thickness, $E_{\mathrm{p}}(\theta)$ (both $\theta$ and
$\theta-1$ will be required: $\theta-1$ to compensate for the
missing dimer trench in the underlying $(2 \times N)$ reconstruction,
and $\theta$ to compensate for the missing rows of dimers in the top
layer), as well as $E_{\mathrm{p}_\mathrm{strip}}(\theta)$.

\subsection{\label{sec:approx}Testing Approximations}
\begin{figure}
\includegraphics[width=\columnwidth,clip]{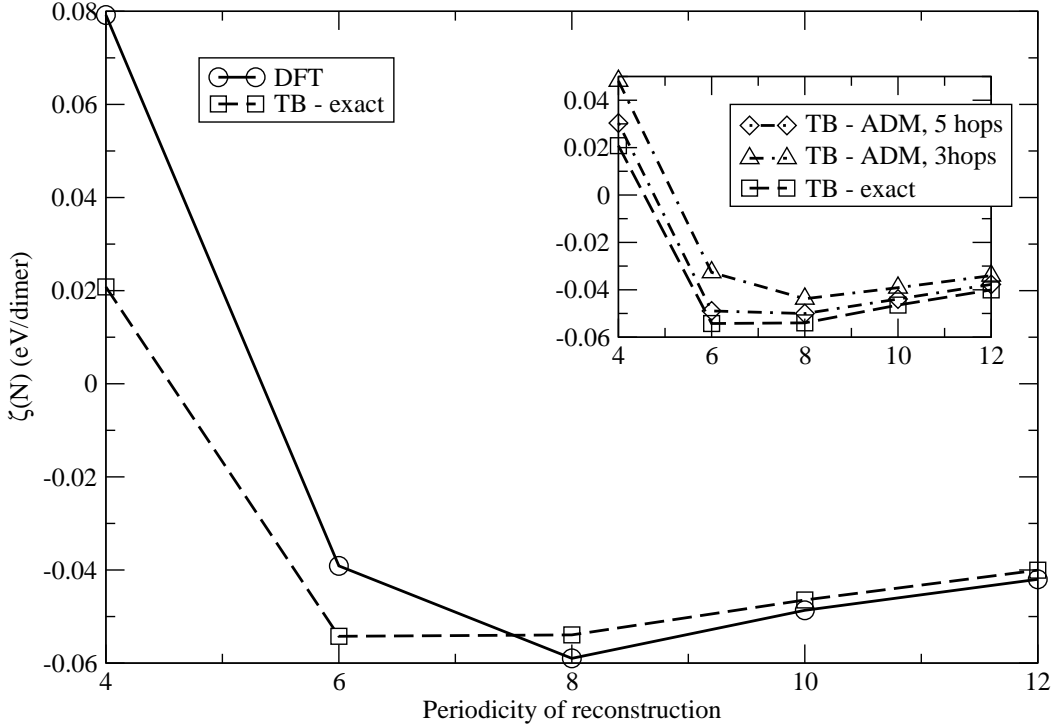}
\caption{\label{fig:CompareMeth1ML}The energy change per Ge dimer to
form a periodically reconstructed surface from a perfect one,
$\zeta(N)$, for the $(2 \times N)$ Ge/Si(001) surface plotted against
\textit{N}, calculated using DFT and exact diagonalisation tight
binding.  Inset: the same quantities calculated using exact
diagonalisation tight binding and ${\mathcal O}(N)$ tight binding
(ADM) with density matrix cutoffs of 5 and 3 hamiltonian-range hops.
See text for a detailed discussion.}
\end{figure}

We now consider the errors introduced by the two approximations that
we use: tight binding, rather than full \textit{ab initio} theory; and
the cutoff on the density matrix used in the linear scaling technique.
In figure~\ref{fig:CompareMeth1ML}, we present the results for
$\zeta(N)$ (defined above in equation~\ref{eq:zeta}) for sub-monolayer
coverages of Ge on Si(001).  The main graph shows the results found
using exact diagonalisation plotted with the DFT results from our
earlier paper\cite{Oviedo2002}, while the inset graph shows the
${\mathcal O}(N)$ results plotted with the exact diagonalisation
results.  The technique used to calculate $\zeta(N)$ is discussed in
detail in Ref.~\cite{Oviedo2002} and briefly above in
Sec.~\ref{sec:Gemu}.

Considering the \textit{ab initio} and tight binding exact
diagonalisation results first, we can see that the agreement is good
for large \textit{N}, but that at $N=4$ in particular there is a
considerable discrepancy.  However, the reason for this is not due to
the approximation made in using tight binding.  The lattice mismatch
of the \textit{ab initio} results\cite{Oviedo2002} is too large
(calculated to be 5.7\% as opposed to the experimental and (fitted)
tight binding value of 4\%) which will increase any strain-related
effects; also, the \textit{ab initio} calculations allowed only the
top four layers of the slab to relax. Using the same conditions as the
\textit{ab initio} calculations (top four layers relaxing, and
rescaling the tight binding parameters to create a 5.7\% lattice
mismatch) we find $\zeta(N=4) = 0.054$ eV, where the \textit{ab
initio} value is 0.079 eV, which is in much better agreement.  This
suggests that the discrepancy between the tight binding and \textit{ab
initio} results for small \textit{N} is in large part due to the
different lattice parameter and simulation conditions, and that the
tight binding parameterisation has introduced only minor inaccuracies
relative to the \textit{ab initio} calculations.

We must also consider the effect of cutting off the density matrix
(the essential approximation in creating an ${\mathcal O}(N)$ method)
on the tight binding results.  Data from the \textsc{DensEl}
calculations for two different density matrix cutoffs (three and five
hamiltonian range hops, which correspond roughly to 6 \AA\ and 10 \AA)
and the exact diagonalisation are shown in the inset to
Fig.~\ref{fig:CompareMeth1ML}.  We can see that, at 5 hops, the
${\mathcal O}(N)$ results and the exact diagonalisation results are in
almost exact agreement, while at 3 hops the overall shape of the curve
is reproduced well, though the exact numerical values are a little
high.  However, these results are sufficiently accurate for our
purposes, and we will use a 3 hop cutoff in all calculations in the
paper.  These results are in excellent agreement with previous
work\cite{Bowler1997} which suggested that energy differences were
completely converged at 5 hops and well converged at 3 hops.

\section{\label{sec:results}Results}

In this section we present results for a series of increasingly complex
reconstructions for coverages of Ge on Si(001) that exceed 1ML: calculations
of the value of \textit{N} for the $(2 \times N)$ reconstruction; 
calculations of the energy of kinks in the missing-dimer trenches for this
reconstruction (for 1ML as well as more than 1ML); and finally the 
$(M \times N)$ ``patch'' reconstruction.

\subsection{\label{sec:2byNDeep}The $(2 \times N)$ reconstruction for 
more than 1ML}

Under certain growth conditions, the $(2 \times N)$
reconstruction is seen to persist when more than one monolayer of Ge has
been deposited.  This results in a more strained layer, and
the appearance of the reconstruction changes: the spacing of the 
missing dimer trenches decreases, and the trenches become straighter.
In this section, we investigate the effect on periodicity, while in 
the next section (Sec.~\ref{sec:kinking}) we model the straightening
of the trenches.

The energy used for comparisons between different periodicities,
$E_p(\theta)$, is $-10.35$ eV/dimer for $\theta=1$ ML, $-10.08$ eV/dimer
for $\theta=2$ ML, and $-10.11$ eV/dimer for $\theta=3$ ML.  The small
increase in this energy from 2ML to 3ML is surprising, but can be
attributed to relaxation effects in the substrate (the third layer of
Ge allows more relaxation than might be expected).

\begin{figure}
\includegraphics[width=\columnwidth,clip]{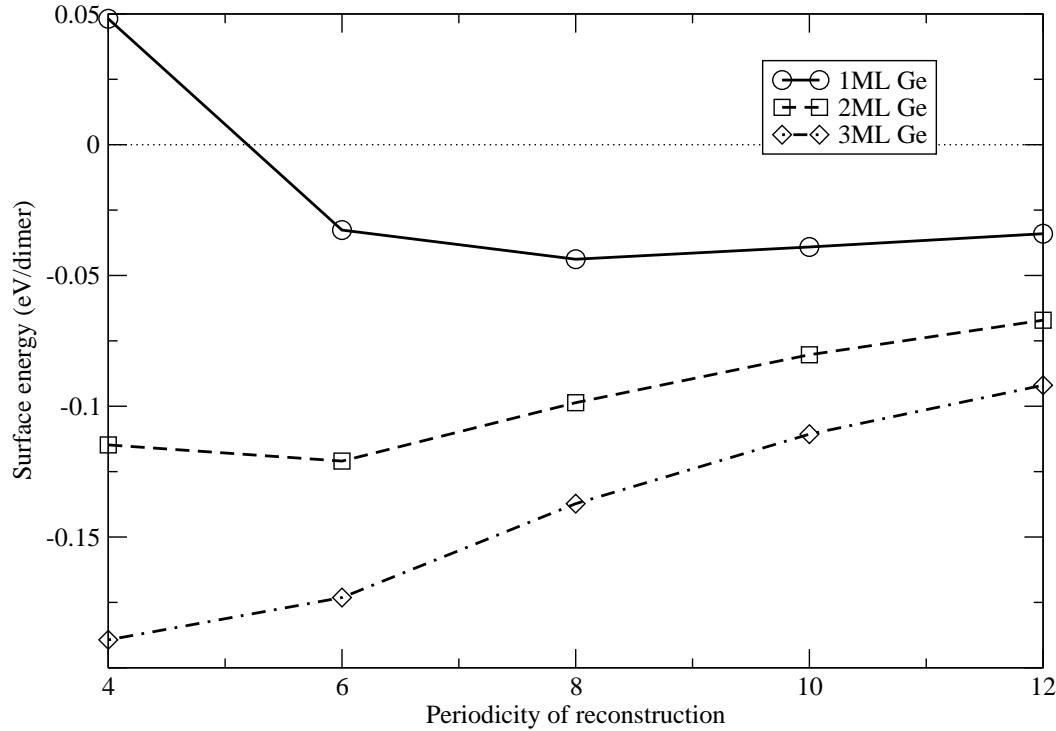}
\caption{\label{fig:2byN2+ML}The surface energy per Ge dimer for the
$(2 \times N)$ Ge/Si(001) surface plotted against \textit{N} for 1, 2 and 3ML
coverages.}
\end{figure}

The results for different Ge ML coverage are given in
table~\ref{tab:2byN2+ML} and Figure~\ref{fig:2byN2+ML}. As shown in
the figure, the optimal \textit{N} value decreases when Ge coverage
increases from 1ML to 2 ML, and again from 2 ML to 3 ML. This result
indicates that increasing the Ge coverage results in more and more
vacancies to release evolving stress and reduce strain
energy. Figure~\ref{fig:2byN2+ML} also shows that the increase in
energy with \textit{N} becomes steeper at the deeper coverages,
indicating that the statistical distibution of \textit{N} would become
narrower with increasing Ge coverage (since the larger values of
\textit{N} become more expensive energetically), as also shown by
STM\cite{Goldfarb1997}.  These results are broadly in agreement with
what would be expected for an increasingly strained surface, and point
towards the next stage in stress relief in this system: three
dimensional islands (often known as ``hut clusters'').  The value of
$N=4$ (or smaller) for 3ML is a little surprising --- it is rather
smaller than seen in experiment, or reported previously in modelling.
The discrepancy with previous modelling is not surprising, as we use a
different method of comparison with experiment.  The discrepancy with
experiment is somewhat more troubling, and will be discussed more
in Section~\ref{sec:conc}.

\begin{table}
\caption{\label{tab:2byN2+ML}Formation energy of $(2 \times N)$
reconstruction for the Ge/Si(100) surface relative to perfect
Ge/Si(100) surface and $\zeta(N)$ for increasing Ge coverage
(described above in Sec.~\protect\ref{sec:Gemu}).  Energies were
obtained using ADM.}
\begin{tabular}{lcccccc} 
\hline
& $\theta_{Ge}$ (ML)  &     4     &    6     &    8     &   10    &    12   \\
\hline						     		
$E_\mathrm{f}$ & 1    &  10.492   & 10.184   & 10.041   &  9.995  &   9.974  \\
\hline 						     		
$\zeta(N)$     & 1    &   0.048   & -0.033   & -0.044   & -0.039  &  -0.034  \\
\hline 						     		
$E_\mathrm{f}$ & 2    &   9.738   &  9.478   &  9.393   &  9.360  &   9.345  \\
\hline	   					     		
$\zeta(N)$     & 2    &  -0.115   & -0.121   & -0.099   & -0.080  &  -0.067  \\
\hline     					     		
$E_\mathrm{f}$ & 3    &   9.544   &  9.246   &  9.151   &  9.116  &   9.101  \\
\hline	   					     		
$\zeta(N)$     & 3    &  -0.189   & -0.173   & -0.137   & -0.111  &  -0.092  \\
\hline       
\end{tabular}
\end{table}

\subsection{\label{sec:kinking}Ordering the $(2 \times N)$ trenches}

\begin{figure}
\begin{tabular}{c|c|c}
\includegraphics[width=0.25\columnwidth,clip]{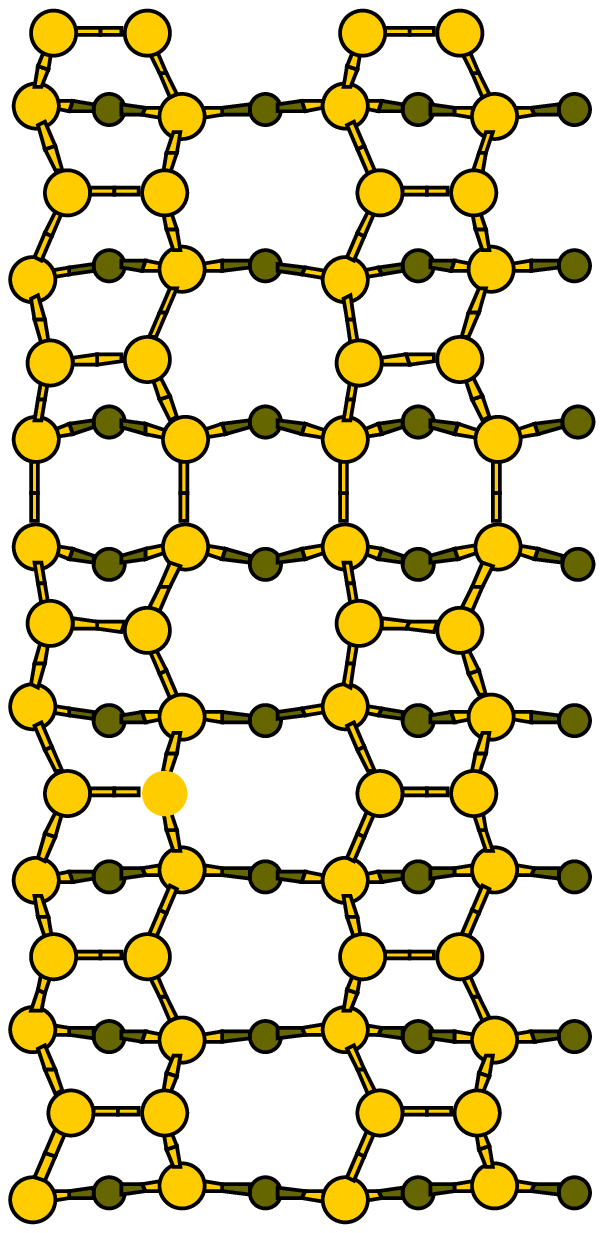} &
\includegraphics[width=0.25\columnwidth,clip]{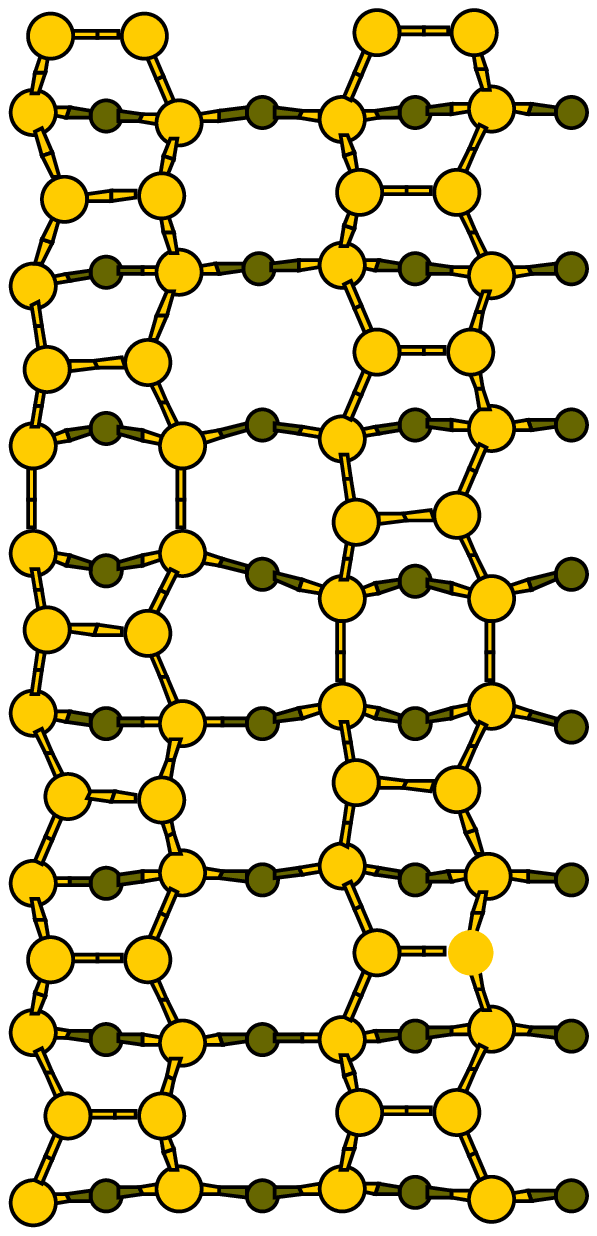} &
\includegraphics[width=0.25\columnwidth,clip]{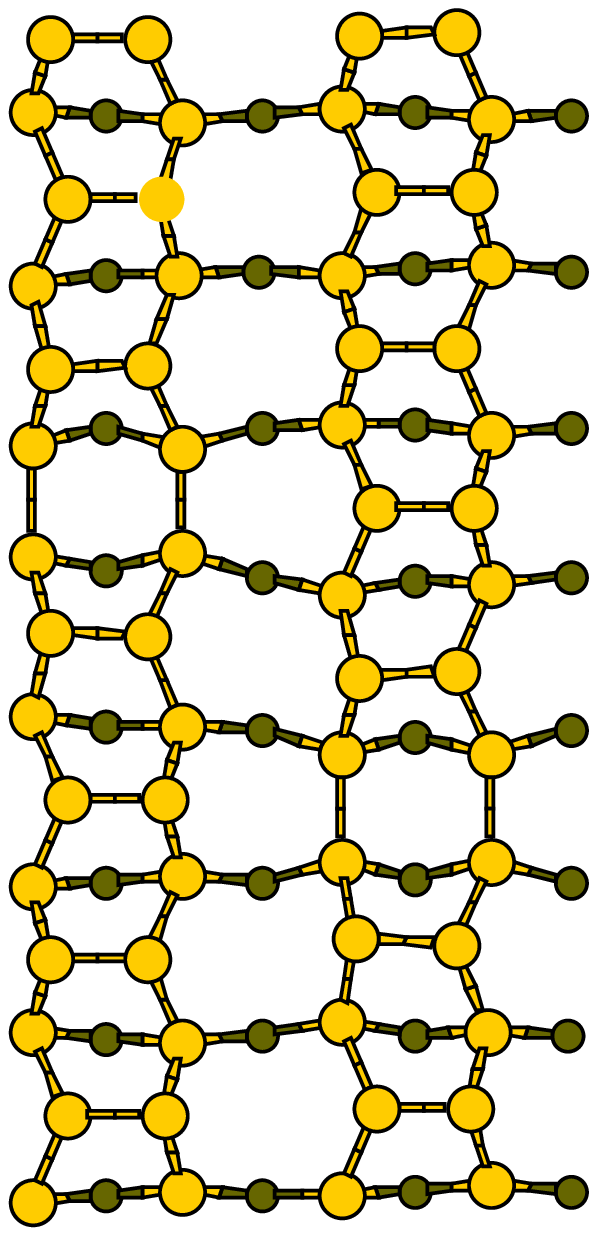} \\
(a) & (b) & (c)
\end{tabular}
\caption{\label{fig:kinking}Kinking in missing dimer trenches for two
monolayers of Ge on Si(001).  Ge is shown as light circles, Si as dark
circles.  (a) A missing dimer trench with no kink. (b) A kink of depth
one in a missing dimer trench (in the right-hand dimer row, the
missing dimer is displaced vertically by one dimer). (c) A kink of
depth two in a missing dimer trench (the missing dimer is displaced
vertically by two dimers)}
\end{figure}

In experiments where the $(2 \times N)$ reconstruction persists with
increasing Ge coverage, as well as gradually reducing the periodicity
\textit{N}, the missing dimers become more ordered --- i.e.  they line
up to form straight trenches.  The obvious explanation for this is that
it becomes energetically more favourable for the missing dimers to align
with increasing Ge depth, and it is this phenomenon that we investigate
here.  Our strategy is to calculate the energy cost of disordering a
trench of missing dimers as a function of the depth of Ge on the surface.
We will introduce isolated kinks into straight trenches (i.e. a 
displacement of one or more dimers as we go from one dimer row to an
adjacent one, as illustrated in Fig.~\ref{fig:kinking}) and calculate
the energy difference between this configuration and a perfect trench.

We assume that only the top layer of Ge is defective, and use the
optimal value of \textit{N} for each system as calculated in
section~\ref{sec:2byNDeep}.  We have performed kinking calculations on
unit cells with two, four and six dimer rows, in order to find the
interaction between adjacent kinks (i.e. in the two dimer row cell,
the kinks will be as close as possible, while in the six dimer row
cell, they will be separated by three dimer rows each).  This effect
has been found to be small.

The trench kinking energy of different depths of Ge coverage (1, 2 and
3ML) is shown in table~\ref{tab:Kink}. To give an idea of how likely
kinks of different depths are, we have also calculated the
distribution of the trench kink depths at 600$^\circ$C, assuming that
these are distributed in thermal equilibrium.  We have take the
probability of a given kink depth, $l$, at any site to be given by a
Boltzmann factor, $P_l \propto e^{E_l/k_B T}$, and have normalised the
distribution to obtain a proportion of depths:
\begin{equation}
\sum_l e^{E_l/k_B T} = 1.
\end{equation}
We can see that at all coverages, a straight trench (kink depth of
zero) is most energetically favorable, becoming more so as the depth
of Ge (and hence strain) increases.  For a depth of 1ML of Ge, 45\% of
trench sites will kink, leading to the highly disordered appearance
seen in STM.  As the coverage increases, the probability of kinking
decreases, so that at 3ML only about 6\% of trench sites will kink by
2, and 72\% of kink sites will remain straight.  It is interesting to
note that at all coverages the energy \textit{difference} from a kink
depth of 1 to 2 is larger than from a kink depth of 0 to 1, suggesting
that there is a strong strain along the trench.  Looking at
Fig.~\ref{fig:kinking}, we can see this strain in the third layer
silicon atoms (shown as dark circles) which are significantly more
twisted in (c) than in (a).  This is orthogonal to the strain that we
have seen previously which was \textit{along} the dimer row.  These
results emphasise the important role that strain plays in the
morphologies of the $(2 \times N)$ reconstructed surface in the
process of growing Ge/Si(001).
   
\begin{table}
\caption{\label{tab:Kink}Trench kinking energy of $(2 \times N)$
reconstructed Si(100) with 1, 2 and 3 Ge ML coverage, and kink
distribution of kink depth. The temperature considered here is 600$^\circ$C.}
\begin{tabular}{ccccc} 
\hline
kink depth  & $\theta_{Ge}$(ML) & Kink energy(eV) & Population  \\
\hline
   0       &       1      &   0.000     &   0.554      \\ 
   1       &       1      &   0.036     &   0.348      \\
   2       &       1      &   0.132     &   0.098      \\  
\hline			  
   0       &       2      &   0.000     &   0.645      \\ 
   1       &       2      &   0.069     &   0.258      \\
   2       &       2      &   0.142     &   0.097      \\   
\hline			  
   0       &       3      &   0.000     &   0.717      \\ 
   1       &       3      &   0.089     &   0.219      \\
   2       &       3      &   0.182     &   0.064      \\ 
\hline  
\end{tabular}
\end{table} 

\subsection{\label{sec:mbyn}The $(M \times N)$ reconstruction}

Having considered the $(2 \times N)$ reconstruction for multiple
layers of Ge, we now move on to the two-dimensional $(M \times N)$
reconstruction (or the ``patch'' reconstruction), for 2 and 3 ML
coverage of Ge, illustrated in Fig.~\ref{fig:IllusMbyN} for 2 ML of
Ge.  For the case of 2 ML of Ge, we fixed \textit{N} at 8 (from our
previous calculation for a single monolayer) and changed the value
\textit{M} to find the most energetically favorable. We used the
method described in Sec.~\ref{sec:Gemu} to calculate $\zeta(M)$ for
the $(8 \times M)$ reconstruction, with
$E_{\mathrm{p}_\mathrm{strip}}(\theta) = -10.03$~eV. These results are
presented in table~\ref{tab:MbyN2ML}. We can see that the optimal
\textit{M} is 6, that is, of $(8 \times M)$ reconstructions, the $(8
\times 6)$ is most energetically favorable and most stable structure.
When modelling the 3 Ge ML coverage case, we used a periodicity of 6
for \textit{N} (again, considering our earlier results for the $(2
\times N)$ surface with 2 Ge ML) and
$E_{\mathrm{p}_\mathrm{strip}}(\theta) = -10.02$~eV.  As shown also in
table~\ref{tab:MbyN2ML}, we found that the optimal \textit{M} for $(6
\times M)$ reconstruction is between 4 and 6.  From these results we
can see that the $(M \times N)$ reconstruction is still an effective
strain relief mechanism, and that for three layers of Ge more relief
is obtained (since the value of $\zeta(M)$ is more negative).
 
\begin{table}
\caption{\label{tab:MbyN2ML}$\zeta(M)$ for the $(M \times N)$
reconstruction at 2 and 3ML of Ge coverage (for a definition of
$\zeta(M)$, see Sec.~\protect\ref{sec:Gemu}).  The value of \textit{N}
was fixed at 8 (2ML) and 6 (3ML) in accordance with results for the
$(2 \times N)$ reconstruction in Sec.~\protect\ref{sec:2byNDeep}.}
\begin{tabular}{cccccc} 
\hline
            & $\theta_{Ge}$ (ML) &   4    &   6    &   8    &   10   \\
\hline
$\zeta(M)$  & 2                & -0.199 & -0.262 & -0.202 & -0.178 \\
\hline  
$\zeta(M)$  & 3                & -0.275 & -0.271 & -0.216 & -0.175 \\
\hline  
\end{tabular}
\end{table} 
  
\subsection{\label{sec:Compare1D2D}Comparing the $(2 \times N)$ and $(M \times N)$ reconstructions}

We have discussed in Sec.~\ref{sec:Gemu} how we will compare the
stabilities of the $(2 \times N)$ and $(M \times N)$ reconstructions
relative to the perfect surface, and these results are given in
table~\ref{tab:Compare1D2D}.  We used values of
$E_{\mathrm{p}}(\theta)$ and $E_{\mathrm{p}_\mathrm{strip}}(\theta)$
given above in Sections~\ref{sec:2byNDeep} and \ref{sec:mbyn} to
calculate these energies.

\begin{table}
\caption{\label{tab:Compare1D2D}Energy per dimer per dimer row
relative to perfect layers for $(2 \times N)$ and $(M \times N)$
reconstructions for 2 and 3 layers of Ge on Si(001), given in
eV/dimer/dimer row.  The final three columns represent the value of
\textit{N} for the $(2 \times N)$ reconstruction and \textit{M} for
the $(M \times N)$ reconstruction.  For the $(M \times N)$
reconstruction, the value of \textit{N} is given in the first column.}
\begin{tabular}{lcccc}
\hline
Reconstruction     & $\theta_{Ge}$ (ML) &   4      &   6       &   8   \\
\hline
$(2 \times N)$       &    2             & -0.0862  &  -0.1008  &  -0.0862  \\
$(M \times N) (N=8)$ &    2             & -0.0048  &  -0.0233  &  -0.0135  \\
\hline
$(2 \times N)$       &    3             & -0.1420  &  -0.1444  &  -0.1201  \\
$(M \times N) (N=6)$ &    3             & -0.0970  &  -0.1062  &  -0.0952  \\
$(M \times N) (N=8)$ &    3             & -0.1220  &  -0.1284  &  -0.1134  \\
\hline
\end{tabular}
\end{table}

We see that the $(2 \times N)$ reconstruction is more stable than the
$(M \times N)$ for a 2ML coverage of Ge, and also for 3ML though by a
much smaller margin.  It is interesting that the stability of the $(M
\times N)$ reconstructions for 3ML is increased when the second layer
has a periodicity of $N=8$ (though the most stable periodicity for the
$(2 \times N)$ reconstruction with two layers of Ge is $N=6$).
Nevertheless, these reconstructions are all stable relative to a
perfect surface, and their formation under different growth conditions
will depend on a subtle interplay of factors, such as: deposition
rate; substrate temperature; and growth source.

\section{\label{sec:conc}Conclusions}

Various important results can be drawn from our modelling of several
layers of Ge on Si(001).  First, we have shown that tight binding, and
in particular ${\mathcal O}(N)$ tight binding, is sufficiently
accurate to be used for modelling complex, strain-driven
reconstructions such as those found in this system.  Second, we have
identified the trends behind the $(2 \times N)$ reconstruction with
increasing Ge depth, and shown that the trenches of missing dimers
will tend to line up and straighten as the depth of Ge deposited
increases.  Third, we have explored the $(M \times N)$ ``patch''
reconstruction for different Ge coverages, contrasting it with the $(2
\times N)$ reconstruction, and shown that, while it is less stable
than the $(2 \times N)$ reconstruction for all coverages of Ge
considered, it becomes less unstable as the depth of Ge deposited
increases.

Our results are based on two fundamental assumptions: full thermal
equilibrium; and a lack of intermixing.  We have already seen one area
where our results appear not to be in exact agreement with experiment
(in the value of \textit{N} for the $(2 \times N)$ reconstruction),
and it is quite possible that these assumptions may be partly
responsible for this discrepancy.  The experimental observations
generally take place on substrates which roughen as growth proceeds,
leading to finite domains, which will have a considerable effect
on the values of \textit{N} observed.  The values of \textit{M} found
for the $(M \times N)$ reconstruction show a similar trend towards
values of $M=4$, though not as sharply (the value of $M=6$ for three
layers of Ge is essentially degenerate with the value of $M=4$).  At
this size of reconstruction, some of the forces governing stability
are due to the elastic constants of the system, which tight binding
may not reproduce exactly.  Nevertheless, our results reproduce and
explain observed trends rather well, and may be considered as a
limiting case, where the simplest assumptions apply: thermal
equilibrium and complete Ge/Si segregation.

\ack 
We thank the UK Engineering and Physical Sciences Research Council and
the Royal Society (DRB) and the Chinese Government (KHL) for funding.
We gratefully acknowledge useful conversations with Jaime Oviedo, Ilan
Goldfarb and James Owen.

\bibliography{MbyN}

\begin{thebibliography}{10}
\expandafter\ifx\csname url\endcsname\relax
  \def\url#1{\texttt{#1}}\fi
\expandafter\ifx\csname urlprefix\endcsname\relax\def\urlprefix{URL }\fi

\bibitem{Mo1991}
Y.-W.Mo, M.G.Lagally, J.\ Crystal Growth 111 (1991) 876.

\bibitem{Oviedo2002}
J.Oviedo, D.R.Bowler, M.J.Gillan, Surf.\ Sci. 515 (2002) 483.

\bibitem{Goldfarb1997}
I.Goldfarb, J.H.G.Owen, P.T.Hayden, D.R.Bowler, K.Miki, G.A.D.Briggs, Surf.\
  Sci. 394 (1997) 105.

\bibitem{Voigt1999}
B.Voigtl{\"a}nder, M.K{\"a}stner, Phys.\ Rev.\ B 60 (1999) R5121.

\bibitem{Wu1995}
F.Wu, X.Chen, Z.Zhang, M.G.Lagally, Phys.\ Rev.\ Lett. 74 (1995) 574.

\bibitem{Huang1997}
K.-H.Huang, T.-S.Ku, D.-S.Lin, Phys.\ Rev.\ B 56 (1997) 4878.

\bibitem{Kohler1992}
U.K{\"o}hler, O.Jusko, B.M{\"u}ller, M.~von Hoegen, M.Pook, Ultramicroscopy
  42-44 (1992) 832.

\bibitem{Tomitori1994}
M.Tomitori, K.Watanabe, M.Kobayashi, O.Nishikawa, Appl.\ Surf.\ Sci. 76-77
  (1994) 322.

\bibitem{Goldfarb1997b}
I.Goldfarb, P.T.Hayden, J.H.G.Owen, G.A.D.Briggs, Phys.\ Rev.\ Lett. 78 (1997)
  3959.

\bibitem{Goedecker1999}
S.Goedecker, Rev.\ Mod.\ Phys. 71 (1999) 1085.

\bibitem{Tersoff1991}
J.Tersoff, Phys.\ Rev.\ B 43 (1991) 9377.

\bibitem{Tersoff1992}
J.Tersoff, Phys.\ Rev.\ B 45 (1992) 8833.

\bibitem{Liu1996}
F.Liu, M.G.Lagally, Phys.\ Rev.\ Lett. 76 (1996) 3156.

\bibitem{Uberuaga2000}
B.P.Uberuaga, M.Leskovar, A.P.Smith, H.J{\'o}nsson, M.Olmstead, Phys.\ Rev.\
  Lett. 84 (2000) 2441.

\bibitem{Goringe1997}
C.M.Goringe, D.R.Bowler, E.H.Hern{\'a}ndez, Rep.\ Prog.\ Phys. 60 (1997) 1447.

\bibitem{Bowler1998}
D.R.Bowler, M.Fearn, C.M.Goringe, A.P.Horsfield, D.G.Pettifor, J.\ Phys.:\
  Condens.\ Matter 10 (1998) 3719.

\bibitem{Bowler2002}
D.R.Bowler, J.\ Phys.:Condens.\ Matt. 14 (2002) 4527.

\bibitem{Goringe1995}
C.M.Goringe, Computational modelling of semiconductor surfaces, Ph.D. thesis,
  Oxford University (1995).
\newline\urlprefix\url{http://www.cmmp.ucl.ac.uk/~drb/DensEl.html}

\bibitem{Li1993}
X.-P.Li, R.W.Nunes, D.Vanderbilt, Phys.\ Rev. B 47 (1993) 10891.

\bibitem{Monkhorst1976}
H.J.Monkhorst, J.D.Pack, Phys.\ Rev.\ B 13 (1976) 5188.

\bibitem{Bowler1997}
D.R.Bowler, M.Aoki, C.M.Goringe, A.P.Horsfield, D.G.Pettifor, Modell.\ Simul.\
  Mat.\ Sci.\ Eng. 5 (1997) 199.

\end{thebibliography}
\end{document}